\documentclass[prl,twocolumn]{revtex4-1}                                                                                                                                                                                                                                                                                                                                                        
\usepackage{amssymb}
\usepackage{empheq}
\usepackage{color}

\usepackage[utf8]{inputenc}  
\usepackage{amsmath}    
\usepackage{amsfonts}
\usepackage{mathrsfs} % $\mathscr{L}$   
\usepackage[unicode]{hyperref}
\usepackage{graphicx}
\newcommand{\beq}{\begin{equation}}
\newcommand{\eeq}{\end{equation}}
\newcommand{\bea}{\begin{eqnarray}}
\newcommand{\eea}{\end{eqnarray}}

\newcommand{\ep}{\epsilon}

\begin{document}

\title{The critical geometry of a thermal Big Bang}
\author{Niayesh Afshordi}
%\email{nafshordi@pitp.ca}
\affiliation{Perimeter Institute for Theoretical Physics, 31 Caroline St. N., Waterloo, ON, N2L 2Y5, Canada}
\affiliation{Department of Physics and Astronomy, University of Waterloo, Waterloo, ON, N2L 3G1, Canada}
\author{Jo\~ao Magueijo}                                                                                              
\affiliation{Theoretical Physics, Blackett Laboratory, Imperial College, London, SW7 2BZ, United Kingdom}

\begin{abstract}
We explore the space of scalar-tensor theories containing two non-conformal
 metrics, and find a discontinuity pointing to a ``critical'' cosmological solution. 
Due to the different maximal speeds of propagation for matter and gravity,
the cosmological fluctuations start off inside the horizon even without inflation, and will more naturally have a thermal origin (since there is never vacuum domination). The critical model makes an unambiguous, non-tuned prediction for the spectral index of the scalar fluctuations:  $n_S= 0.96478(64)$. Considering also that no gravitational waves are produced, we have unveiled the most predictive model on offer. The model has a simple geometrical interpretation 
as a probe 3-brane embedded in an $EAdS_2\times E_3$ geometry.
\end{abstract}

\date{\today}    

\maketitle

{\it 1. Introduction.}
In spite of its mathematical simplicity and observational triumphs, the Big Bang model of the Universe  
remains an unfinished work of art. 
%Upon closer scrutiny, m
Many of its late-time successes can be traced to the initial conditions {\it postulated} for its early stages, and these are put in by hand, without justification, other than to retrofit the data.  The main culprit for this shortcoming is the so-called horizon problem: the cosmological structures we observe today span scales that lay outside the ever-shrinking ``horizons'' of physical contact that plagued the early universe. This precludes a causal explanation for their initial conditions. 

Several extensions of the Big Bang model have been proposed with the aim of opening up its horizons.  An early bout of accelerated expansion~\cite{Guth,Linde,Albrecht}, a contracting phase followed by a bounce~\cite{Turok}, a loitering early stage~\cite{Nayeri}, and a varying speed of light  (VSL)~\cite{Moffat,AlMag} have all been considered. None of these proposals evades the criticism that retrofitting the data is still used to select  {\it in detail} the primordial fluctuations that the model {\it should} produce. 
Once primordial causal contact is established, work can start on concrete physical mechanisms 
for spoiling perfect homogeneity (e.g. vacuum quantum fluctuations or thermal fluctuations).
Typically it is found that one can produce a wide range of initial conditions including, but not circumscribed to those explaining the 
observations. 

Specifically, the primordial density fluctuations can be characterised by an amplitude $A_S$, measuring their intensity at a given scale,  and a spectral index $n_S$, measuring how the amplitude varies with scale. Observations~\cite{Ade:2015xua,Ade:2015lrj}
show that $A_S =2.142(49) \times 10^{-9}$ and 
$n_S=0.9667(40)$, signalling a very specific slightly red spectrum, i.e. one with enhanced amplitude for longer wavelengths. Whilst the 
observed $A_S$ probably indicates nothing more than a hierarchy between two energy scales, one might have expected 
a theoretical forecast for $n_S$. Yet, all theories effectively adjust their free parameters 
(e.g. the reheating temperature after inflation, or the number of e-foldings) to fit the observed $n_S$, 
from within a range of possibilities. This is
not to say that they entirely lack predictivity; indeed they do predict a plethora of conditions involving $n_S$ and other 
observables (e.g.~\cite{piazza}).

In this {\it Letter}, we revisit a class of VSL models~\cite{Magueijo:2003gj} in which there are two non-conformal metrics, one for matter and another for gravity, so that light and other massless matter particles travel faster than gravity. Conditions for the observational success of such models have been identified~\cite{Magueijo:2008pm,Clayton:1999zs, Magueijo:2008sx,piazza,Kinneyetal,Agarwal:2014ona},
considering both a vacuum and a thermal initial state. In common with other models, they do not bypass the criticism voiced above. However, in this {\it Letter} we uncover a remarkable result pertaining to thermal scenarios.

It is known that thermal VSL models require a fast phase transition in $c$ so as to produce near-scale-invariant fluctuations;
however, the scale-invariant limit ($n_S=1$) is unreachable. Closer inspection of the space of all possible theories reveals that this is due to a discontinuity, pointing to a special, critical solution that should be regarded as the preferential model for a phase transition in $c$.
Instead of $n_S=1$, the thermal fluctuations in this model display a running $n_S<1$. But what is truly notable is that the model has a single free parameter, so that the amplitude $A_S$ fully fixes the value of $n_S$ at the observationally relevant scales. The predicted value is within current constraints, but improved observations would unambiguously prove or rule out the theory. 
The model also has a simple geometrical interpretation as a probe 3-brane embedded in an $EAdS_2\times E_3$ geometry.

{\it 2. The critical model.}
We start by reviewing the general framework of scalar-tensor bimetric theories. 
In these models there are two metrics (or frames): $g_{\mu\nu}$ associated with the gravitational action (the Einstein frame), and $\hat g _{\mu\nu}$, to which matter is minimally coupled (the matter frame).  
The action takes the general form:
\beq
S=\frac{M_P^2}{2} \int d^4 x\sqrt{-g}R[g_{\mu\nu}]+
\int d^4 x\sqrt{-\hat g}{\cal L}_M(\Psi,{\hat g}_{\mu\nu}) + 
S_\phi \nonumber \eeq
where $M_P$ is the reduced Planck mass, and $S_\phi$ encodes the dynamics of the field $\phi$ relating the two metrics. If the metrics are conformally related we have a  ``varying-$G$''  theory, such as Brans-Dicke theory. In ``varying-$c$'' theories, rather, the metrics are non-conformally related:
\beq\label{metric}
{\hat g}_{\mu\nu}=g_{\mu\nu}+B(\partial_\mu \phi)(\partial_\nu\phi),
\eeq
so that the light cones spanned by massless matter particles and by gravitons do not coincide~\cite{Clayton:1999zs,Magueijo:2008sx}. 
In general $B$ (also known as the {\it warp factor}, for reasons to be made obvious soon) is a function of $\phi$.
If the speed of light is to be larger than that of gravity, then $B>0$ (with
signature $+ - - - $). 

It may seem that the number of theories of this type is endless, but this is not the case.
The simplest non-trivial  $S_\phi$ must consist of two generally non-constant cosmological terms, 
one in the matter frame and the other in the Einstein frame: 
\beq\label{Sphi}
S_\phi=\int d^4 x\sqrt{-\hat g}(-2\Lambda_m(\phi)) +  \int d^4
x\sqrt{-g}(-2 \Lambda_g(\phi))\; .\eeq
Furthermore, only one of the functions $\Lambda_m(\phi)$, $\Lambda_g(\phi)$ and $B(\phi)$ is independent \footnote{While these functions, in general, can depend $(\partial\phi)^2$ and higher derivatives, we assume that such dependence is suppressed by a UV scale (e.g., Planck energy) and can be neglected in the regime of validity of effective field theory.}. This has been known
for a while; here we sketch the proof in two steps.

Firstly, let $\phi$ be canonically normalized in the generalized sense that it should exhibit a Klein-Gordon equation of motion in the matter frame when no matter is present. Then, its action should be a cosmological term in the matter 
frame, due to a simple calculation in variational calculus in the presence of two metrics (see~\cite{Clayton:1999zs,Magueijo:2008sx}). 
Furthermore, if the field
dynamics is to be driven by $2\Lambda_m$, regarded as a potential, we should have 
$2\Lambda_m(\phi)=-1/B(\phi)$. This fully fixes the first term of (\ref{Sphi}) in terms of $B$, and it is known~\cite{Magueijo:2008sx} that it amounts to postulating a
DBI action in the Einstein frame
\beq\label{DBI}
S_\phi=\int d^4 x\sqrt{- g}\left(\frac{1}{B}\sqrt{1+2BX} - 
V\right)\; ,\eeq
with $X=\frac{1}{2}(\partial_\mu\phi)(\partial^\mu\phi)$ and $V=2\Lambda_g$. This can be derived from simple properties of
determinants, and will be important in seeking a geometrical interpretation for our critical model.

Secondly, of the two remaining free functions ($B$ and $V$) only one is free in the UV limit, which is the limit of interest 
to us. In the physical situation we are considering, the action (\ref{DBI}) should more strictly be called 
anti-DBI, since the sign of $B$ is opposite to the usual one, so that the speed of light is larger, rather than smaller than that of gravity. 
Thus, the UV limit of the theory is achieved with $X\gg 1$ (instead of saturating at an upper bound, as is the case with the usual DBI theory), so that:
\beq\label{cusculag}
{\cal L}_\phi\approx \sqrt\frac{2X}{B}- V +{\cal O} \left(1 \over \sqrt{B^3 X} \right).
\eeq
This is nothing but the cuscuton model~\cite{Afshordi:2006ad,Afshordi:2007yx}, and indeed the speed of sound
is infinite in this limit ($c_s \approx \sqrt{2 B X} \rightarrow \infty$). The model has conformal (Weyl) symmetry~\cite{Afshordi:2009tt,Horava:2009uw}, 
so that any scale-factor
$a(t)$ is a solution. This implies that spatial flatness is compulsory and fully fixes $V$~\cite{Afshordi:2006ad,Afshordi:2007yx,Afshordi:2009tt}. If $\rho$ and $p$ denote density and pressure,  we have $\rho\approx V$ and $p+\rho\approx K$~\cite{GarrigaandMukhanov,Afshordi:2006ad}, where $K=\dot \phi/\sqrt B$ is the kinetic energy. The (spatially flat) Friedmann and continuity equations are:
\beq\label{bianchicus}
3M^2_PH^2\approx V ~~{\rm and}~~
%\dot \rho + 3H (p+\rho)=
\dot V+ 3 H K\approx 0, 
\eeq
where $H=\dot a/a$.
% (non-flat models contradict the Bianchi identities and conformal invariance of the model). 
These can be integrated as:
\beq\label{Vgsln}
V(\phi)=\frac{3}{4 M^2_P}\left(\int\frac{d\phi}{\sqrt{B(\phi)}}\right)^2 +{\cal O}\left(\epsilon V \over c^2_s \right)\; ,
\eeq
fixing $V$ as a function of $B$.
%and
%\beq\label{htphi}
%H=\frac{1}{2}\int\frac{d\phi}{\sqrt B}.
%\eeq
Here $\epsilon=-\dot H/H^2=\frac{3}{2}(1+w)$, where $w=p/\rho$, and $c_s$ is the speed of sound in the matter frame.
Although any $w$ is possible,  it can be shown that as $c^2_s = \partial p/\partial\rho|_X \rightarrow\infty$, 
the next order corrections yield $w =p/\rho \rightarrow \infty$ for generic solutions~\cite{SM,prep}.

So far we have merely reviewed old results. Now we come to the crucial element of this {\it Letter}. It has been shown~\cite{Magueijo:2008pm,piazza,Agarwal:2014ona} that 
thermal bimetric scenarios are close to scale-invariance whenever $B(\phi)\propto \phi^n $, with $n$ close to 2. Then, the potential
$V$ is still a power-law, but its exponent is close to zero (cf. Eqn.~(\ref{Vgsln})). The variation in $c_s$ is abrupt, but one still has constant
$\epsilon_s=\dot c_s/(c_s H)$, with $\epsilon_s\rightarrow -\infty$ as  $n\rightarrow 2$. The cosmological solutions are 
``scaling solutions'', i.e. they have constant $\epsilon$ and 
$\epsilon_s$, leading to thermal fluctuations with constant $n_S$,
which can be tuned to be as close to 1 as wanted.
Indeed 
\beq\label{nthermal}
n_S-1=\frac{\ep+1}{\ep_s+\ep-1}\; , \eeq
and although $n_S=1$ is unreachable, any red spectrum as close to scale-invariance as required can be obtained by
suitably tuning $B$.

It should be immediately obvious from Eqn.~(\ref{Vgsln}) the reason why the scale-invariant limit cannot be reached. 
Within the space of these theories, there is a discontinuity at 
$B\propto \phi^2$, because the potential fails to be a power-law. 
All the theories around it imply power-law potentials, but this ``critical''  theory stands out as an exception:
\bea
B_{\rm crit.}(\phi)&=&B_0\left(\frac{\phi}{M_P}\right)^2,\label{critical1}\\
\Rightarrow V_{\rm crit.}(\phi)&=&\frac{3}{4 B_0}  \ln^2\left(\frac{\phi}{M_P}\right) \label{critical2}.
\eea
It marks a special, crucial boundary in the space of theories. 
The critical model is unique in that it cannot have 
a constant $\epsilon_s$, since $V$ is no longer a power-law. This induces natural deviations
from scale-invariance, making its phenomenology remarkable, as we will show presently.

{\it 3. Geometrical interpretation.}
Before embarking upon the phenomenology of the critical model we reinforce its special status by 
uncovering an elegant geometrical interpretation. It is known that the DBI action can be derived from the induced metric on a 
probe 3-brane embedded in a higher dimensional geometry, with the $B$ function interpreted as a geometrical ``warp'' factor. 
For example, in the celebrated DBI action associated with the motion of a probe 3-brane in $AdS_5 \times S_5$  geometry,  
one finds  $B \propto -\phi^{-4}$, with interesting cosmological implications~\cite{Maldacena:1997re,Silverstein:2003hf}.  

Likewise, $B(\phi) \propto \phi^2$ follows from embedding a 3-brane in the $EAdS_2 \times E_3$ geometry given by:
\beq
d\tau_5^2= \frac{r^2}{R^2}dt^2+\frac{R^2}{r^2}dr^2 -dx^2-dy^2-dz^2,
\eeq
where $R$ is the radius of the Euclidean $AdS_2$. 
Ignoring the gravitational backreaction, the induced action on a uniform probe 3-brane at $r(t)$ is given by
\beq\label{S3B}
S_{3B} = T_3  \int d^4x \sqrt{\frac{r^2}{R^2}+ \frac{R^2\dot{r}^2}{r^2}},
\eeq
where $T_3$ is the brane tension (with mass units $M^4$).  Field $\phi$ is a redefinition of $r$ that 
renders (\ref{S3B}) canonical in the IR limit, and a Taylor expansion shows that this is given by $ r=4R^3 T_3 / \phi^2$. 
Straightforward algebra shows that this brings (\ref{S3B})  to the 
anti-DBI form  (\ref{DBI}) with $B$ matching the critical model (\ref{critical1}) and 
\beq
B_0 = \left(M_P \over 2 R  T_3  \right)^2.
\eeq
Turning on gravity for the effective 4D geometry, the potential is fixed by 
%demanding consistency of continuity and Friedmann equations in the cuscuton limit, according to 
Eq. (\ref{Vgsln}), as a result of the conformal invariance of the theory in the UV~\cite{Afshordi:2009tt,Horava:2009uw}.

A crucial novelty here is that the extra dimension, $r$, is time-like rather than space-like, something also discussed in string theory literature~\cite{Dijkgraaf:2016lym}. While this may raise alarm about ghost instabilities for the bulk,  the ghost degrees of freedom may be made arbitrarily heavy and thus decouple from the 4D low-energy effective field theory~\cite{prep}.

{\it 4. Density fluctuations.}
We now come to the core of this {\it Letter}, the evaluation of the thermally induced fluctuations for the critical solution. 
This can be done following well-known methods developed for theories with a varying speed of sound $c_s$~\cite{GarrigaandMukhanov,Magueijo:2008sx,Agarwal:2014ona}, since that is what our theory is in the Einstein frame.
The second order action for the curvature fluctuation $\zeta$ is:
 \beq
{\cal  S}_2= \frac{1}{2} M^2_P \int d\eta d^3 {\rm x} ~z^2 \left[ \zeta'^2 - c_s^2 (\nabla \zeta)^2 \right],
 \eeq
where $z=\frac{a}{c_s}\sqrt{2\epsilon}$ and $\eta$ is conformal time. Therefore we have a standard quantum field theory in variable $v= M_P z\zeta$, subject to dynamical equation:
\beq\label{veq}
v'' +{\left (c_s^2 k^2-\frac{z''}{z}\right)}v=0,
\eeq
where $k$ is the comoving wave-number. 
The central quantity to be computed is $c_s$, and this is given by~\cite{GarrigaandMukhanov,Afshordi:2006ad,Magueijo:2008sx}:
\beq\label{sound_crit}
c_s =  \sqrt{1+ 2BX}\approx \frac{2}{3}\epsilon B \rho \approx 
\frac{2}{3}\epsilon B_0\rho e^{ 4\sqrt{ \frac{B_0 \rho}{3}}},
\eeq
where the first identity is generic for (anti-)DBI models, in the second step we  used 
$\frac{2}{3}\epsilon=\frac{K}{V}\approx \frac{K}{\rho}$ and $K\approx \sqrt{2X/B}$, and in the third we used
Eqns.~(\ref{critical1}) and~(\ref{critical2}). The fact that $V$ (and so $\rho$) is not a power-law in $\phi$, 
explains why the model has a $c_s$ which is not a power-law in $a$ or $\rho$. 
Even if the background scales (constant $\epsilon$), the speed of sound does not, with a varying $\epsilon_s$ given by:
\beq
\epsilon_s=\frac{\dot c_s}{c_s H}=-2\epsilon{\left(1+2\sqrt{\frac{B_0\rho}{3}}\right)}.
\eeq
Thus, $n_S$ is expected to run, a property that can be guessed from (\ref{nthermal}). However that formula is 
incorrect for varying $\epsilon_s$, indeed many standard formulae in the literature~\cite{GarrigaandMukhanov,Magueijo:2008sx,Agarwal:2014ona} break down. A full derivation of $n_S$ can be found in the Supplementary Material~\cite{SM} (SM) (in this version of our paper included in appendix).
Here we present an approximate calculation, good enough to extract all the salient features. 

As usual, Eq.~(\ref{veq}) has two regimes, an acoustic one and a gravitational instability one, depending on which of its two terms in $v$ dominates. 
The two regimes are separated by the sound horizon scale, $k_h$, where these terms become of the same order:
\beq\label{kh}
c_s^2k_h^2\sim \frac{z''}{z}\approx (aH\epsilon_s)^2-(aH\epsilon_s)' \approx (aH\epsilon_s)^2,
\eeq
(for simplicity, we have assumed a constant $\epsilon$, but in fact this is not necessary). The sound horizon scale therefore satisfies $c_s k_h\approx aH\epsilon_s$, and we note the extra factor of $\epsilon_s$ with regards to the usual formula. Matching the 2 types of solution is sufficient to derive to a good approximation the power spectrum frozen-in 
outside the horizon. For $k\gg k_h$ the solutions should be normalized as~\cite{GarrigaandMukhanov,Magueijo:2008sx,Agarwal:2014ona}:
\beq
v=\frac{e^{-i\int c_s k\, d\eta}}{\sqrt{2c_s k}},
\eeq
whereas for $k\ll k_h$ the growing mode takes the form $v=F(k)z$. By means of simple algebra $F(k)$ can be found by matching the two expressions at $k\sim k_h$. 

The square of $F(k)$ is nothing but the frozen-in power spectrum of $\zeta$, up to a factor representing the expectation value of $2{\hat N}+1$, where ${\hat N}$ is the number operator (note that upon quantization $v^2$ is multiplied by ${\hat a}^\dagger \hat a + \hat a \hat a^\dagger$, where $\hat a$ is an annihilation operator). 
For vacuum fluctuations, this factor is simply 1, whereas for a thermal state it is twice the thermal occupation number of mode $k$ in the Rayleigh-Jeans limit~\cite{Magueijo:2008sx,Agarwal:2014ona}. This is $2 T_c/k$, where $T_c=Tc_s/a$ is a ``conformal temperature'' which remains constant during the varying-$c$ phase~\cite{Magueijo:2008sx,Agarwal:2014ona}. 
The frozen-in dimensionless power spectrum of the thermally induced fluctuations is therefore:
\beq\label{calP}
{\cal P}^{th}_\zeta (k)\equiv\frac{k^3}{2\pi^2}{\langle |\zeta|^2\rangle}^{th}\approx 
\frac{1}{24\pi^2}\frac{\epsilon _s^2}{\epsilon}\frac{\rho}{c_sM_p^4}\frac{T_c}{k},
\eeq
where the right hand side is to be evaluated at horizon crossing (now  $c_s k \approx aH\epsilon_s$).
We stress the extra factors in $\epsilon_s$ found in Eq.~(\ref{kh}) and (\ref{calP}), in relation to standard formulae~\cite{GarrigaandMukhanov}. They are irrelevant if $\epsilon_s$ is a constant, but not in our case. 
Eq.~(\ref{calP}) is valid up to factors of order one (fully restored in the SM~\cite{SM}; see appendix). 

%We note at once that the model has a single free parameter, the 4-volume scale $B_0$. 

Combining Eqs.~(\ref{kh}) and (\ref{calP}) and using the chain rule we find for the spectral index:
\bea
n_S-1= \frac{d\ln {\cal P}_\zeta^{th} }{d \ln k} 
= -\frac{1+2\ep}{4\ep} \left(B_0 \rho \over 3\right)^{-\frac{1}{2}} + {\cal O} \left(\frac{1}{B_0\rho}\right),\nonumber
\label{ns_model}
\eea
%and for its running:
%\beq
%\frac{dn_s}{d \ln k} = -\frac{3}{16\ep}\frac{(1+2\ep)}{B_0 \rho} + {\cal O}\left(B_0 \rho\right)^{-3/2}, \nonumber\label{run_model}
%\eeq
where $\rho$ is the density when $k=k_h$. 
Thus, $n_S$ runs from very red, at the largest scales, to almost scale-invariant, at the smallest. However, 
as announced, the observed amplitude $A_S$ fixes where we are in this running flow. 
Note that the model has a single free parameter, the 4-volume scale $B_0$.
% and 
%$B_0\rho$ is fixed by the amplitude and therefore
%$n_s$ is fully determined. 
Although Eq~(\ref{calP})
seems to depend both on $B_0 M_P^4$ and $B_0\rho$, the former can be eliminated by 
using $c_s k=aH \epsilon_s$, the Friedman equation, and some basic thermodynamics, to recast it in the descriptive form:
\beq\label{eqbrho}
{\cal P}^{th}_\zeta (k)\frac{g_0T_0^3}{k^3}=C  (B_0\rho)^{1+\frac{2}{\ep}} \exp\left(4\sqrt{3B_0\rho}\right)
\eeq
where $C$ is a numerical constant~\cite{SM}. The left hand side can be evaluated from observations. For a given mode (say, $k = 0.05~{\rm Mpc}^{-1}$) the first factor is the observed amplitude (${\cal P}_\zeta (k)=2.142(49) \times
10^{-9} $, \cite{Ade:2015xua}), and the second is the dimensionless entropy inside the scale $k$ nowadays (with $g_0=3.91$ the effective number of relativistic degrees of freedom). As stated above (and in~\cite{SM}), generically $\epsilon\rightarrow \infty$, so we can solve (\ref{eqbrho}) to get $B_0\rho \approx 583.03(16)$, where the uncertainties are both observational and arise from the fact that the model is only reliable to 
$ {\cal O} (\frac{1}{B_0\rho})$. 
%(and to a much less extent from observational uncertainties)] \textcolor{red}{[NA: For reasons not entirely clear to me, the error in $B_0\rho$ is mainly due to observational error on amplitude of power spectrum, while the error in $n_s$ is due to $ {\cal O} (\frac{1}{B_0\rho})$ effects. ]} .  
Using (\ref{ns_model}) we thus obtain:
\beq 
n_S= 0.96478(64),
\eeq
well within the most stringent current observational constraints (viz. $n_S=0.9667 (40)$, cf.~\cite{Ade:2015xua}). 

This is a remarkable result. 
But the model makes further predictions. It produces no tensor modes (since the horizon problem is not solved for gravitons), and so singles out a point in the $\{n_S, r\}$ diagram, with $r=0$. 
It also predicts  (cf. Eqs.~(\ref{kh}),  (\ref{calP}) and (\ref{ns_model})) the running of the spectral index to be: 
\beq
\frac{dn_S}{d\ln k} = -\frac{3}{2} (n_s-1)^2 \approx  - 1.8 \times 10^{-3},
\eeq
within the allowed observational range of $(-6.5 \pm 7.6) \times 10^{-3}$ (see~\cite{Ade:2015xua}). 
As for non-Gaussianity we find an amplitude for the bispectrum of order unity, $f_{NL} = {\cal O}(1)$, comparable to similar models~\cite{piazza} but with a very different and unique shape (to be reported elsewhere~\cite{prep}). We have unveiled the most predictive model on offer.  

What are the provisos of our claims? From the above we can work out that 
$B_0 M_P^4\approx 6.6 \times 10^{13}$, so that the energy scale at the end of the transition is
$ 3.5\times 10^{-4} M_P$, with the current horizon scale leaving the sound horizon less than 3 orders of magnitude above this. 
So we never exceed the Planck scale (in common with other thermal varying-$c$ scenarios~\cite{Magueijo:2008sx,Agarwal:2014ona}),
allaying the first obvious criticism.  Then, there are model uncertainties. 
The equation of state can have an effect on the final result (for example, $\epsilon=2$ would push $n_S$ down to 
$n_S=0.95292$); however we have arguments for why
$\epsilon\gg 1$ is generic in our model(see~\cite{SM} and Appendix). Furthermore,  
in evaluating $T_c$ at horizon crossing we have assumed entropy conservation in the constant $c$ phase
(the change in $g$ drops out of the final result), but more importantly we have assumed no ``reheating'' at the end of the 
varying-$c$ phase. This is because in our scenario any such process would be {\it ad hoc} and unnecessary, since the universe 
is always hot. Nonetheless, we note that a reheat by a factor of, for example,  $10^{10}$  would push $n_S$ up to $n_S=0.96838$. 
An isothermal gluing of the two phases remains the most minimal assumption.

{\it 5. Discussion.}
In summary, we built upon previous work on thermal fluctuations in bimetric scenarios which showed that a sufficiently fast phase transition in $c_s$ leads to fluctuations as close to scale-invariance as seen in the data~\cite{Magueijo:2008sx,Agarwal:2014ona}. In such scenarios, fitting the observed $n_S$  requires fine-tuning the warp factor $B(\phi)$. Here we improved on this by discovering that the reason why exact scale-invariance is never achieved is that the limit is discontinuous, pointing to a critical solution with quadratic warp factor,
but a non-power-law potential (fully determined by the Bianchi identities and UV conformal symmetry). The critical solution has a simple geometrical interpretation as the (anti-)DBI action of a probe 3-brane embedded in an $EAdS_2\times E_3$ geometry. The non-power-law nature of the potential induces a non-scaling speed of sound, which in turn produces a natural red tilt and running of the power spectrum.

But what makes the model remarkable is that 
the amplitude $A_S$ for a given scale fixes its location on this overarching structure, leading to a single prediction for the observed $n_S$. The model does not require reheating, and this is the ultimate reason why it is more predictive than inflation, even if factors external to cosmology were to pre-select one of inflation's many models. Inflationary models invariably predict a range of  $n_S$, depending on the number of e-foldings, or the reheating temperature (even for a fixed choice of inflaton action). Adding to this the fact that our model makes precise predictions for the level of {\it primordial} gravitational waves ($r=0$), the running of $n_S$, and non-Gaussianity, we can conclude without prejudice that we have in hand a very predictive model indeed. The fact that its main prediction (for $n_S$) lies spot in the middle of the Planck results should not beguile us into a false sense of security. Improved observations will soon vindicate or disprove this model. 

One may wonder about the status in our model of the other cosmological problems, such as the flatness, homogeneity and isotropy problems. Firstly, the view may be held that such historically motivating problems are now considered to be of lesser importance than explaining the structure of our Universe, or may even be misguided~\cite{carroll}. Nonetheless we remark that it is possible to solve them using the VSL mechanism {\it before} the phase transition (e.g.~\cite{Clayton:1999zs,AlMag}). In other scenarios it may also happen that their solution takes place in a different phase to structure formation. Furthermore, we find that at least the flatness problem can be solved, in a single package, during the phase transition. The conformal symmetry of the theory in the UV~\cite{Afshordi:2009tt,Horava:2009uw} not only fixes the potential but requires exact flatness (Eqs.(\ref{bianchicus}) lead to a contradiction in the presence of spatial curvature).  A full investigation of these matters is deferred to~\cite{prep}.

{\it Acknowledgments}
We would like to thank Robert Brandenberger, Keith Copsey, Giulia Gubitosi and Sarah Shandera for helpful discussions. 
The work of NA was partially supported by the Natural Science and Engineering Research Council of Canada, the University of Waterloo and by Perimeter Institute for Theoretical Physics. Research at Perimeter Institute is supported by the Government of Canada through Industry Canada and by the Province of Ontario through the Ministry of Research \& Innovation. 
JM acknowledges support from the John Templeton Foundation, an STFC consolidated grant and the
Leverhulme Trust, and thanks the Perimeter Institute for hospitality.

\bibliography{Bibliography.bib}

\begin{thebibliography}{29}%
\makeatletter
\providecommand \@ifxundefined [1]{%
 \@ifx{#1\undefined}
}%
\providecommand \@ifnum [1]{%
 \ifnum #1\expandafter \@firstoftwo
 \else \expandafter \@secondoftwo
 \fi
}%
\providecommand \@ifx [1]{%
 \ifx #1\expandafter \@firstoftwo
 \else \expandafter \@secondoftwo
 \fi
}%
\providecommand \natexlab [1]{#1}%
\providecommand \enquote  [1]{``#1''}%
\providecommand \bibnamefont  [1]{#1}%
\providecommand \bibfnamefont [1]{#1}%
\providecommand \citenamefont [1]{#1}%
\providecommand \href@noop [0]{\@secondoftwo}%
\providecommand \href [0]{\begingroup \@sanitize@url \@href}%
\providecommand \@href[1]{\@@startlink{#1}\@@href}%
\providecommand \@@href[1]{\endgroup#1\@@endlink}%
\providecommand \@sanitize@url [0]{\catcode `\\12\catcode `\$12\catcode
  `\&12\catcode `\#12\catcode `\^12\catcode `\_12\catcode `\%12\relax}%
\providecommand \@@startlink[1]{}%
\providecommand \@@endlink[0]{}%
\providecommand \url  [0]{\begingroup\@sanitize@url \@url }%
\providecommand \@url [1]{\endgroup\@href {#1}{\urlprefix }}%
\providecommand \urlprefix  [0]{URL }%
\providecommand \Eprint [0]{\href }%
\providecommand \doibase [0]{http://dx.doi.org/}%
\providecommand \selectlanguage [0]{\@gobble}%
\providecommand \bibinfo  [0]{\@secondoftwo}%
\providecommand \bibfield  [0]{\@secondoftwo}%
\providecommand \translation [1]{[#1]}%
\providecommand \BibitemOpen [0]{}%
\providecommand \bibitemStop [0]{}%
\providecommand \bibitemNoStop [0]{.\EOS\space}%
\providecommand \EOS [0]{\spacefactor3000\relax}%
\providecommand \BibitemShut  [1]{\csname bibitem#1\endcsname}%
\let\auto@bib@innerbib\@empty
%</preamble>
\bibitem [{\citenamefont {Guth}(1981)}]{Guth}%
  \BibitemOpen
  \bibfield  {author} {\bibinfo {author} {\bibfnamefont {A.~H.}\ \bibnamefont
  {Guth}},\ }\href {\doibase 10.1103/PhysRevD.23.347} {\bibfield  {journal}
  {\bibinfo  {journal} {Phys. Rev.}\ }\textbf {\bibinfo {volume} {D23}},\
  \bibinfo {pages} {347} (\bibinfo {year} {1981})}\BibitemShut {NoStop}%
%%CITATION = PHRVA,D23,347;%%
\bibitem [{\citenamefont {Linde}(1982)}]{Linde}%
  \BibitemOpen
  \bibfield  {author} {\bibinfo {author} {\bibfnamefont {A.~D.}\ \bibnamefont
  {Linde}},\ }\bibfield  {booktitle} {\emph {\bibinfo {booktitle} {{Second
  Seminar on Quantum Gravity Moscow, USSR, October 13-15, 1981}}},\ }\href
  {\doibase 10.1016/0370-2693(82)91219-9} {\bibfield  {journal} {\bibinfo
  {journal} {Phys. Lett.}\ }\textbf {\bibinfo {volume} {B108}},\ \bibinfo
  {pages} {389} (\bibinfo {year} {1982})}\BibitemShut {NoStop}%
%%CITATION = PHLTA,B108,389;%%
\bibitem [{\citenamefont {Albrecht}\ and\ \citenamefont
  {Steinhardt}(1982)}]{Albrecht}%
  \BibitemOpen
  \bibfield  {author} {\bibinfo {author} {\bibfnamefont {A.}~\bibnamefont
  {Albrecht}}\ and\ \bibinfo {author} {\bibfnamefont {P.~J.}\ \bibnamefont
  {Steinhardt}},\ }\href {\doibase 10.1103/PhysRevLett.48.1220} {\bibfield
  {journal} {\bibinfo  {journal} {Phys. Rev. Lett.}\ }\textbf {\bibinfo
  {volume} {48}},\ \bibinfo {pages} {1220} (\bibinfo {year}
  {1982})}\BibitemShut {NoStop}%
%%CITATION = PRLTA,48,1220;%%
\bibitem [{\citenamefont {Steinhardt}\ and\ \citenamefont
  {Turok}(2002)}]{Turok}%
  \BibitemOpen
  \bibfield  {author} {\bibinfo {author} {\bibfnamefont {P.~J.}\ \bibnamefont
  {Steinhardt}}\ and\ \bibinfo {author} {\bibfnamefont {N.}~\bibnamefont
  {Turok}},\ }\href {\doibase 10.1126/science.1070462} {\bibfield  {journal}
  {\bibinfo  {journal} {Science}\ }\textbf {\bibinfo {volume} {296}},\ \bibinfo
  {pages} {1436} (\bibinfo {year} {2002})}\BibitemShut {NoStop}%
%%CITATION = SCIEA,296,1436;%%
\bibitem [{\citenamefont {Nayeri}\ \emph {et~al.}(2006)\citenamefont {Nayeri},
  \citenamefont {Brandenberger},\ and\ \citenamefont {Vafa}}]{Nayeri}%
  \BibitemOpen
  \bibfield  {author} {\bibinfo {author} {\bibfnamefont {A.}~\bibnamefont
  {Nayeri}}, \bibinfo {author} {\bibfnamefont {R.~H.}\ \bibnamefont
  {Brandenberger}}, \ and\ \bibinfo {author} {\bibfnamefont {C.}~\bibnamefont
  {Vafa}},\ }\href {\doibase 10.1103/PhysRevLett.97.021302} {\bibfield
  {journal} {\bibinfo  {journal} {Phys. Rev. Lett.}\ }\textbf {\bibinfo
  {volume} {97}},\ \bibinfo {pages} {021302} (\bibinfo {year} {2006})},\
  \Eprint {http://arxiv.org/abs/hep-th/0511140} {arXiv:hep-th/0511140 [hep-th]}
  \BibitemShut {NoStop}%
%%CITATION = HEP-TH/0511140;%%
\bibitem [{\citenamefont {Moffat}(1993)}]{Moffat}%
  \BibitemOpen
  \bibfield  {author} {\bibinfo {author} {\bibfnamefont {J.~W.}\ \bibnamefont
  {Moffat}},\ }\href {\doibase 10.1142/S0218271893000246} {\bibfield  {journal}
  {\bibinfo  {journal} {Int. J. Mod. Phys.}\ }\textbf {\bibinfo {volume}
  {D2}},\ \bibinfo {pages} {351} (\bibinfo {year} {1993})},\ \Eprint
  {http://arxiv.org/abs/gr-qc/9211020} {arXiv:gr-qc/9211020 [gr-qc]}
  \BibitemShut {NoStop}%
%%CITATION = GR-QC/9211020;%%
\bibitem [{\citenamefont {Albrecht}\ and\ \citenamefont
  {Magueijo}(1999)}]{AlMag}%
  \BibitemOpen
  \bibfield  {author} {\bibinfo {author} {\bibfnamefont {A.}~\bibnamefont
  {Albrecht}}\ and\ \bibinfo {author} {\bibfnamefont {J.}~\bibnamefont
  {Magueijo}},\ }\href {\doibase 10.1103/PhysRevD.59.043516} {\bibfield
  {journal} {\bibinfo  {journal} {Phys. Rev.}\ }\textbf {\bibinfo {volume}
  {D59}},\ \bibinfo {pages} {043516} (\bibinfo {year} {1999})},\ \Eprint
  {http://arxiv.org/abs/astro-ph/9811018} {arXiv:astro-ph/9811018 [astro-ph]}
  \BibitemShut {NoStop}%
%%CITATION = ASTRO-PH/9811018;%%
\bibitem [{\citenamefont {Ade}\ \emph {et~al.}(2015{\natexlab{a}})\citenamefont
  {Ade} \emph {et~al.}}]{Ade:2015xua}%
  \BibitemOpen
  \bibfield  {author} {\bibinfo {author} {\bibfnamefont {P.~A.~R.}\
  \bibnamefont {Ade}} \emph {et~al.} (\bibinfo {collaboration} {Planck}),\
  }\href@noop {} {\  (\bibinfo {year} {2015}{\natexlab{a}})},\ \Eprint
  {http://arxiv.org/abs/1502.01589} {arXiv:1502.01589 [astro-ph.CO]}
  \BibitemShut {NoStop}%
%%CITATION = ARXIV:1502.01589;%%
\bibitem [{\citenamefont {Ade}\ \emph {et~al.}(2015{\natexlab{b}})\citenamefont
  {Ade} \emph {et~al.}}]{Ade:2015lrj}%
  \BibitemOpen
  \bibfield  {author} {\bibinfo {author} {\bibfnamefont {P.~A.~R.}\
  \bibnamefont {Ade}} \emph {et~al.} (\bibinfo {collaboration} {Planck}),\
  }\href@noop {} {\  (\bibinfo {year} {2015}{\natexlab{b}})},\ \Eprint
  {http://arxiv.org/abs/1502.02114} {arXiv:1502.02114 [astro-ph.CO]}
  \BibitemShut {NoStop}%
%%CITATION = ARXIV:1502.02114;%%
\bibitem [{\citenamefont {Magueijo}\ \emph {et~al.}(2010)\citenamefont
  {Magueijo}, \citenamefont {Noller},\ and\ \citenamefont {Piazza}}]{piazza}%
  \BibitemOpen
  \bibfield  {author} {\bibinfo {author} {\bibfnamefont {J.}~\bibnamefont
  {Magueijo}}, \bibinfo {author} {\bibfnamefont {J.}~\bibnamefont {Noller}}, \
  and\ \bibinfo {author} {\bibfnamefont {F.}~\bibnamefont {Piazza}},\ }\href
  {\doibase 10.1103/PhysRevD.82.043521} {\bibfield  {journal} {\bibinfo
  {journal} {Phys. Rev.}\ }\textbf {\bibinfo {volume} {D82}},\ \bibinfo {pages}
  {043521} (\bibinfo {year} {2010})},\ \Eprint {http://arxiv.org/abs/1006.3216}
  {arXiv:1006.3216 [astro-ph.CO]} \BibitemShut {NoStop}%
%%CITATION = ARXIV:1006.3216;%%
\bibitem [{\citenamefont {Magueijo}(2003)}]{Magueijo:2003gj}%
  \BibitemOpen
  \bibfield  {author} {\bibinfo {author} {\bibfnamefont {J.}~\bibnamefont
  {Magueijo}},\ }\href {\doibase 10.1088/0034-4885/66/11/R04} {\bibfield
  {journal} {\bibinfo  {journal} {Rept. Prog. Phys.}\ }\textbf {\bibinfo
  {volume} {66}},\ \bibinfo {pages} {2025} (\bibinfo {year} {2003})},\ \Eprint
  {http://arxiv.org/abs/astro-ph/0305457} {arXiv:astro-ph/0305457 [astro-ph]}
  \BibitemShut {NoStop}%
%%CITATION = ASTRO-PH/0305457;%%
\bibitem [{\citenamefont {Magueijo}(2008)}]{Magueijo:2008pm}%
  \BibitemOpen
  \bibfield  {author} {\bibinfo {author} {\bibfnamefont {J.}~\bibnamefont
  {Magueijo}},\ }\href {\doibase 10.1103/PhysRevLett.100.231302} {\bibfield
  {journal} {\bibinfo  {journal} {Phys.Rev.Lett.}\ }\textbf {\bibinfo {volume}
  {100}},\ \bibinfo {pages} {231302} (\bibinfo {year} {2008})},\ \Eprint
  {http://arxiv.org/abs/0803.0859} {arXiv:0803.0859 [astro-ph]} \BibitemShut
  {NoStop}%
%%CITATION = ARXIV:0803.0859;%%
\bibitem [{\citenamefont {Clayton}\ and\ \citenamefont
  {Moffat}(2000)}]{Clayton:1999zs}%
  \BibitemOpen
  \bibfield  {author} {\bibinfo {author} {\bibfnamefont {M.~A.}\ \bibnamefont
  {Clayton}}\ and\ \bibinfo {author} {\bibfnamefont {J.~W.}\ \bibnamefont
  {Moffat}},\ }\href {\doibase 10.1016/S0370-2693(00)00192-1} {\bibfield
  {journal} {\bibinfo  {journal} {Phys. Lett.}\ }\textbf {\bibinfo {volume}
  {B477}},\ \bibinfo {pages} {269} (\bibinfo {year} {2000})},\ \Eprint
  {http://arxiv.org/abs/gr-qc/9910112} {arXiv:gr-qc/9910112 [gr-qc]}
  \BibitemShut {NoStop}%
%%CITATION = GR-QC/9910112;%%
\bibitem [{\citenamefont {Magueijo}(2009)}]{Magueijo:2008sx}%
  \BibitemOpen
  \bibfield  {author} {\bibinfo {author} {\bibfnamefont {J.}~\bibnamefont
  {Magueijo}},\ }\href {\doibase 10.1103/PhysRevD.79.043525} {\bibfield
  {journal} {\bibinfo  {journal} {Phys. Rev.}\ }\textbf {\bibinfo {volume}
  {D79}},\ \bibinfo {pages} {043525} (\bibinfo {year} {2009})},\ \Eprint
  {http://arxiv.org/abs/0807.1689} {arXiv:0807.1689 [gr-qc]} \BibitemShut
  {NoStop}%
%%CITATION = ARXIV:0807.1689;%%
\bibitem [{\citenamefont {Bessada}\ \emph {et~al.}(2010)\citenamefont
  {Bessada}, \citenamefont {Kinney}, \citenamefont {Stojkovic},\ and\
  \citenamefont {Wang}}]{Kinneyetal}%
  \BibitemOpen
  \bibfield  {author} {\bibinfo {author} {\bibfnamefont {D.}~\bibnamefont
  {Bessada}}, \bibinfo {author} {\bibfnamefont {W.~H.}\ \bibnamefont {Kinney}},
  \bibinfo {author} {\bibfnamefont {D.}~\bibnamefont {Stojkovic}}, \ and\
  \bibinfo {author} {\bibfnamefont {J.}~\bibnamefont {Wang}},\ }\href {\doibase
  10.1103/PhysRevD.81.043510} {\bibfield  {journal} {\bibinfo  {journal}
  {Phys.Rev.}\ }\textbf {\bibinfo {volume} {D81}},\ \bibinfo {pages} {043510}
  (\bibinfo {year} {2010})},\ \Eprint {http://arxiv.org/abs/0908.3898}
  {arXiv:0908.3898 [astro-ph.CO]} \BibitemShut {NoStop}%
%%CITATION = ARXIV:0908.3898;%%
\bibitem [{\citenamefont {Agarwal}\ and\ \citenamefont
  {Afshordi}(2014)}]{Agarwal:2014ona}%
  \BibitemOpen
  \bibfield  {author} {\bibinfo {author} {\bibfnamefont {A.}~\bibnamefont
  {Agarwal}}\ and\ \bibinfo {author} {\bibfnamefont {N.}~\bibnamefont
  {Afshordi}},\ }\href {\doibase 10.1103/PhysRevD.90.043528} {\bibfield
  {journal} {\bibinfo  {journal} {Phys. Rev.}\ }\textbf {\bibinfo {volume}
  {D90}},\ \bibinfo {pages} {043528} (\bibinfo {year} {2014})},\ \Eprint
  {http://arxiv.org/abs/1406.0575} {arXiv:1406.0575 [astro-ph.CO]} \BibitemShut
  {NoStop}%
%%CITATION = ARXIV:1406.0575;%%
\bibitem [{Note1()}]{Note1}%
  \BibitemOpen
  \bibinfo {note} {While these functions, in general, can depend $(\partial
  \phi )^2$ and higher derivatives, we assume that such dependence is
  suppressed by a UV scale (e.g., Planck energy) and can be neglected in the
  regime of validity of effective field theory.}\BibitemShut {Stop}%
\bibitem [{\citenamefont {Afshordi}\ \emph
  {et~al.}(2007{\natexlab{a}})\citenamefont {Afshordi}, \citenamefont {Chung},\
  and\ \citenamefont {Geshnizjani}}]{Afshordi:2006ad}%
  \BibitemOpen
  \bibfield  {author} {\bibinfo {author} {\bibfnamefont {N.}~\bibnamefont
  {Afshordi}}, \bibinfo {author} {\bibfnamefont {D.~J.~H.}\ \bibnamefont
  {Chung}}, \ and\ \bibinfo {author} {\bibfnamefont {G.}~\bibnamefont
  {Geshnizjani}},\ }\href {\doibase 10.1103/PhysRevD.75.083513} {\bibfield
  {journal} {\bibinfo  {journal} {Phys. Rev.}\ }\textbf {\bibinfo {volume}
  {D75}},\ \bibinfo {pages} {083513} (\bibinfo {year} {2007}{\natexlab{a}})},\
  \Eprint {http://arxiv.org/abs/hep-th/0609150} {arXiv:hep-th/0609150 [hep-th]}
  \BibitemShut {NoStop}%
%%CITATION = HEP-TH/0609150;%%
\bibitem [{\citenamefont {Afshordi}\ \emph
  {et~al.}(2007{\natexlab{b}})\citenamefont {Afshordi}, \citenamefont {Chung},
  \citenamefont {Doran},\ and\ \citenamefont {Geshnizjani}}]{Afshordi:2007yx}%
  \BibitemOpen
  \bibfield  {author} {\bibinfo {author} {\bibfnamefont {N.}~\bibnamefont
  {Afshordi}}, \bibinfo {author} {\bibfnamefont {D.~J.~H.}\ \bibnamefont
  {Chung}}, \bibinfo {author} {\bibfnamefont {M.}~\bibnamefont {Doran}}, \ and\
  \bibinfo {author} {\bibfnamefont {G.}~\bibnamefont {Geshnizjani}},\ }\href
  {\doibase 10.1103/PhysRevD.75.123509} {\bibfield  {journal} {\bibinfo
  {journal} {Phys. Rev.}\ }\textbf {\bibinfo {volume} {D75}},\ \bibinfo {pages}
  {123509} (\bibinfo {year} {2007}{\natexlab{b}})},\ \Eprint
  {http://arxiv.org/abs/astro-ph/0702002} {arXiv:astro-ph/0702002 [astro-ph]}
  \BibitemShut {NoStop}%
%%CITATION = ASTRO-PH/0702002;%%
\bibitem [{\citenamefont {Afshordi}(2009)}]{Afshordi:2009tt}%
  \BibitemOpen
  \bibfield  {author} {\bibinfo {author} {\bibfnamefont {N.}~\bibnamefont
  {Afshordi}},\ }\href {\doibase 10.1103/PhysRevD.80.081502} {\bibfield
  {journal} {\bibinfo  {journal} {Phys. Rev.}\ }\textbf {\bibinfo {volume}
  {D80}},\ \bibinfo {pages} {081502} (\bibinfo {year} {2009})},\ \Eprint
  {http://arxiv.org/abs/0907.5201} {arXiv:0907.5201 [hep-th]} \BibitemShut
  {NoStop}%
%%CITATION = ARXIV:0907.5201;%%
\bibitem [{\citenamefont {Horava}(2009)}]{Horava:2009uw}%
  \BibitemOpen
  \bibfield  {author} {\bibinfo {author} {\bibfnamefont {P.}~\bibnamefont
  {Horava}},\ }\href {\doibase 10.1103/PhysRevD.79.084008} {\bibfield
  {journal} {\bibinfo  {journal} {Phys.Rev.}\ }\textbf {\bibinfo {volume}
  {D79}},\ \bibinfo {pages} {084008} (\bibinfo {year} {2009})},\ \Eprint
  {http://arxiv.org/abs/0901.3775} {arXiv:0901.3775 [hep-th]} \BibitemShut
  {NoStop}%
%%CITATION = ARXIV:0901.3775;%%
\bibitem [{\citenamefont {Garriga}\ and\ \citenamefont
  {Mukhanov}(1999)}]{GarrigaandMukhanov}%
  \BibitemOpen
  \bibfield  {author} {\bibinfo {author} {\bibfnamefont {J.}~\bibnamefont
  {Garriga}}\ and\ \bibinfo {author} {\bibfnamefont {V.~F.}\ \bibnamefont
  {Mukhanov}},\ }\href {\doibase 10.1016/S0370-2693(99)00602-4} {\bibfield
  {journal} {\bibinfo  {journal} {Phys.Lett.}\ }\textbf {\bibinfo {volume}
  {B458}},\ \bibinfo {pages} {219} (\bibinfo {year} {1999})},\ \Eprint
  {http://arxiv.org/abs/hep-th/9904176} {arXiv:hep-th/9904176 [hep-th]}
  \BibitemShut {NoStop}%
%%CITATION = HEP-TH/9904176;%%
\bibitem [{SM()}]{SM}%
  \BibitemOpen
  \href@noop {} {\bibinfo  {journal} {See Supplemental Material at [web site]}\
  }\BibitemShut {NoStop}%
\bibitem [{\citenamefont {Afshordi}\ \emph {et~al.}()\citenamefont {Afshordi},
  \citenamefont {Magueijo},\ and\ \citenamefont {Noller}}]{prep}%
  \BibitemOpen
\bibfield  {journal} {  }\bibfield  {author} {\bibinfo {author} {\bibfnamefont
  {N.}~\bibnamefont {Afshordi}}, \bibinfo {author} {\bibfnamefont
  {J.}~\bibnamefont {Magueijo}}, \ and\ \bibinfo {author} {\bibfnamefont
  {J.}~\bibnamefont {Noller}},\ }\href@noop {} {\bibinfo  {journal} {in
  preparation}\ }\BibitemShut {NoStop}%
\bibitem [{\citenamefont {Maldacena}(1999)}]{Maldacena:1997re}%
  \BibitemOpen
\bibfield  {journal} {  }\bibfield  {author} {\bibinfo {author} {\bibfnamefont
  {J.~M.}\ \bibnamefont {Maldacena}},\ }\href {\doibase
  10.1023/A:1026654312961} {\bibfield  {journal} {\bibinfo  {journal} {Int. J.
  Theor. Phys.}\ }\textbf {\bibinfo {volume} {38}},\ \bibinfo {pages} {1113}
  (\bibinfo {year} {1999})},\ \bibinfo {note} {[Adv. Theor. Math.
  Phys.2,231(1998)]},\ \Eprint {http://arxiv.org/abs/hep-th/9711200}
  {arXiv:hep-th/9711200 [hep-th]} \BibitemShut {NoStop}%
%%CITATION = HEP-TH/9711200;%%
\bibitem [{\citenamefont {Silverstein}\ and\ \citenamefont
  {Tong}(2004)}]{Silverstein:2003hf}%
  \BibitemOpen
  \bibfield  {author} {\bibinfo {author} {\bibfnamefont {E.}~\bibnamefont
  {Silverstein}}\ and\ \bibinfo {author} {\bibfnamefont {D.}~\bibnamefont
  {Tong}},\ }\href {\doibase 10.1103/PhysRevD.70.103505} {\bibfield  {journal}
  {\bibinfo  {journal} {Phys. Rev.}\ }\textbf {\bibinfo {volume} {D70}},\
  \bibinfo {pages} {103505} (\bibinfo {year} {2004})},\ \Eprint
  {http://arxiv.org/abs/hep-th/0310221} {arXiv:hep-th/0310221 [hep-th]}
  \BibitemShut {NoStop}%
%%CITATION = HEP-TH/0310221;%%
\bibitem [{\citenamefont {Dijkgraaf}\ \emph {et~al.}(2016)\citenamefont
  {Dijkgraaf}, \citenamefont {Heidenreich}, \citenamefont {Jefferson},\ and\
  \citenamefont {Vafa}}]{Dijkgraaf:2016lym}%
  \BibitemOpen
  \bibfield  {author} {\bibinfo {author} {\bibfnamefont {R.}~\bibnamefont
  {Dijkgraaf}}, \bibinfo {author} {\bibfnamefont {B.}~\bibnamefont
  {Heidenreich}}, \bibinfo {author} {\bibfnamefont {P.}~\bibnamefont
  {Jefferson}}, \ and\ \bibinfo {author} {\bibfnamefont {C.}~\bibnamefont
  {Vafa}},\ }\href@noop {} {\  (\bibinfo {year} {2016})},\ \Eprint
  {http://arxiv.org/abs/1603.05665} {arXiv:1603.05665 [hep-th]} \BibitemShut
  {NoStop}%
%%CITATION = ARXIV:1603.05665;%%
\bibitem [{\citenamefont {Carroll}(2014)}]{carroll}%
  \BibitemOpen
  \bibfield  {author} {\bibinfo {author} {\bibfnamefont {S.~M.}\ \bibnamefont
  {Carroll}},\ }\href@noop {} {\  (\bibinfo {year} {2014})},\ \Eprint
  {http://arxiv.org/abs/1406.3057} {arXiv:1406.3057 [astro-ph.CO]} \BibitemShut
  {NoStop}%
%%CITATION = ARXIV:1406.3057;%%
\bibitem [{\citenamefont {Mukhanov}\ \emph {et~al.}(1992)\citenamefont
  {Mukhanov}, \citenamefont {Feldman},\ and\ \citenamefont
  {Brandenberger}}]{Mukhanov:1990me}%
  \BibitemOpen
  \bibfield  {author} {\bibinfo {author} {\bibfnamefont {V.~F.}\ \bibnamefont
  {Mukhanov}}, \bibinfo {author} {\bibfnamefont {H.}~\bibnamefont {Feldman}}, \
  and\ \bibinfo {author} {\bibfnamefont {R.~H.}\ \bibnamefont
  {Brandenberger}},\ }\href {\doibase 10.1016/0370-1573(92)90044-Z} {\bibfield
  {journal} {\bibinfo  {journal} {Phys.Rept.}\ }\textbf {\bibinfo {volume}
  {215}},\ \bibinfo {pages} {203} (\bibinfo {year} {1992})}\BibitemShut
  {NoStop}%
%%CITATION = PRPLC,215,203;%%
\end{thebibliography}%

\section*{Appendix: Supplemental Material}

\subsection{A. Cosmological Perturbation Theory}

The FRW metric with scalar linear perturbations in the longitudinal gauge is given by:
\beq
ds^2 = a^2(\eta) \left[(1+2\Phi)d\eta^2 -(1-2\Psi)  d {\bf x} \cdot d{\bf x}\right]. 
\eeq
Observational constraints on scalar adiabatic perturbations are often described in terms of the gauge-invariant Bardeen variable \cite{Mukhanov:1990me}:
\beq
\zeta \equiv \Psi -\frac{H}{\dot{H}} (H\Phi +\dot{\Psi}). 
\eeq

Following \cite{GarrigaandMukhanov}, we shall adopt the following quadratic action for the Bardeen variable:
 \beq
{\cal  S}= \frac{1}{2} M^2_P \int dy d^3 {\rm x} ~q^2 \left[ \zeta'^2 - (\nabla \zeta)^2 \right],
 \eeq
 for acoustic waves of speed $c_s$, where
 \bea
 q &\equiv& \frac{a \sqrt{2\epsilon}}{\sqrt{c_s}} ,\label{q_def} \\
 y  &\equiv& \int \frac{c_s dt}{a} = \int c_s d\eta  \label{y_def} ,\\
 ' &\equiv & \frac{\partial}{\partial y},  ~\epsilon  \equiv  - \frac{\dot{H}}{H^2}, \label{eps_def} \\
 M_P  &\equiv&  (8\pi G_N)^{-1/2} = 2.435 \times 10^{18} ~{\rm GeV}.
\eea 
We shall call $y$ the {\it tachyo-conformal} time (which is also equal to the comoving sound horizon), and $M_P$ is the reduced Planck mass. 

We can change to the Mukhanov-Sasaki variable:
\beq
v \equiv M_P q \zeta,
\eeq
 which is canonically normalized:
\beq
 {\cal S}= \frac{1}{2} \int dy d^3 {\rm x}  \left[v'^2 - (\nabla v)^2  + \frac{q''}{q} v^2 \right],
 \eeq
leading to the mode functions that obey the field equation in the Fourier space:               
\begin{equation}
v''_k + \left( k^2 - \frac{q''}{q} \right) v_k = 0.
\end{equation}
%Here the primes represent the derivatives with repect to $y$, i.e. $\frac{d}{dy}$ and $y = \int{c_s d\eta}$ 
%where $\eta$ is conformal time. In accordance with the Speedy Sound Model we can take $q = Ay^{\alpha}$.
%Then $q' = A \alpha y^{\alpha - 1}$ and $q'' = A \alpha (\alpha - 1) y^{\alpha - 2}$.                  
If we have

\beq
q = Q (-y)^{1/2-\nu} \Rightarrow \frac{q''}{q} = \frac{\nu^2 -1/4}{y^2}, \label{q-Q}
\eeq 
 our mode equation becomes

\begin{equation}
v''_k + \left[ k^2 - \frac{\nu^2 -1/4}{y^2} \right] v_k = 0. \label{mode_beta}
\end{equation}

So far, we have only considered the classical equations for linear perturbations. Following the standard canonical quantization procedure, we can decompose the free quantum fields in the Heisenberg picture as:

\begin{equation}
\hat{v}({\bf x},y) = \int{\frac{d^3 {\rm k}}{(2 \pi)^3} \left[ v_k(y) {\hat{a}}_{\bf k} e^{i {\bf k\cdot x} }
 + v_{k}^*(y) {\hat{a}}_{\bf k}^{\dagger} e^{-i {\bf k\cdot x}} \right]},  
\end{equation} 
where ${\hat{a}}_{\bf k}$ and $ {\hat{a}}_{\bf k}^{\dagger}$ are the creation and annihilation operators for particles (or phonons) of momentum ${\bf k}$ around a gaussian vacuum state $\left|0\right\rangle$, which, by definition, has zero particles. 

The adiabatic vacuum state $\left|0\right\rangle_{\rm ad.}$ is defined by the condition that mode functions $v_{k}(y)$ approach the positive frequency (flat space) limit, when $y\rightarrow -\infty$, which is also where adiabatic approximation in Eq. (\ref{mode_beta}) becomes exact:
\beq
v_{k}(y) \rightarrow \frac{\exp(-i ky)}{\sqrt{2k}}, {\rm ~when}~ y \rightarrow -\infty, \label{adiabatic}
\eeq
while its subsequent evolution follows from exactly solving the mode equation (\ref{mode_beta}). This ensures that the adiabatic vacuum coincides with the ground state of the Hamiltonian at infinite past. 

It turns out that Eq. (\ref{mode_beta}) with the initial condition (\ref{adiabatic}) has an exact solution in terms of the Hankel  function of the 2nd kind (or Bessel functions of 1st and 2nd kind):
\beq
v_{k}(y) = \frac{\sqrt{-\pi y}}{2} e^{-i\gamma_\nu}H^{(2)}_\nu (ky),\label{hankel}
\eeq
where 
\beq
 \gamma_\nu = \frac{\pi}{4}(2\nu+1). 
\eeq

The late-time power spectrum of $\zeta$ in a thermal state of temperature $T_*$ is given by:
\begin{equation}
\langle {\cal P}_{\zeta}(k)\rangle_{T_*} = \lim_{y \rightarrow 0^-} \frac{k^3}{2 {\pi}^{2}} \frac{{\left| v_k \right|}^2}{q^2M^2_P} [2 \langle n_k\rangle_{T_*} + 1].\label{power}
\end{equation}
Here, the thermal particle occupation number is given by the Bose-Einstein distribution:
\beq
\langle n_k\rangle_{T_*} = \frac{1}{\exp\left(kc_s\over aT\right)-1}.\label{Bose-Einstein}
\eeq
We can also use the asymptotic form of Hankel function for small arguments:
\beq
\left| v_k \right|^2 = \frac{4^{\nu-1}\Gamma(\nu)^2}{\pi y^{2\nu-1} k^{2\nu}} + {\cal O} (y^{2-2\nu}). \label{asymptote}
\eeq

Combining Eqs. (\ref{q-Q}) with (\ref{power})-(\ref{asymptote}) yields:
\beq
 \langle {\cal P}_{\zeta}(k)\rangle_{T} =\frac{\Gamma[\nu]^2}{\pi^3 M_P^2Q^2} \left[ \frac{2}{\exp\left(kc_s\over aT\right)-1} +1\right] (k/2)^{3-2\nu}.\label{power_wide}
\eeq

Notice that, in the Rayleigh-Jeans limit $kc_s \ll aT$, we have  $\langle n_k\rangle_{T} \propto k^{-1}$, and thus the scalar spectral index is given by:
\beq
n_s-1= 2-2\nu \approx 0 \Rightarrow \nu \approx 1.
\eeq
Therefore, close to scale-invariance in the tachyacoustic phase, the power spectrum takes the form:
\bea
 \langle {\cal P}_{\zeta}(k)\rangle_{T} \approx \frac{T_c}{\pi^3M^2_P Q^2} \approx -\frac{T_c}{\pi^3M^2_P}\frac{dq^{-2}}{dy} \nonumber\\ 
 = \left[-\frac{d\ln(c_s/\ep)}{d\ln a} +2\right]\frac{HT_c}{2\pi^3\ep M^2_Pa },
\eea
where we used the definitions of $q$ and $Q$ (Eq. \ref{q_def} and \ref{q-Q}), while $T_c \equiv aT/c_s$. Furthermore, using the definition $\ep$ (Eq. \ref{eps_def}), and for a rapidly decaying speed of sound $-\ep_s \equiv -\frac{d\ln c_s}{d\ln a} \gg 1$, we can further simplify the expression for the power spectrum:
\beq
  \langle {\cal P}_{\zeta}(k)\rangle_{T} \approx  \frac{d\ln(c_s/\ep)}{d H^{-1}} \frac{T_c}{2\pi^3 M^2_Pa }. \label{power_final}
\eeq
The comoving wavenumber at which the power spectrum freezes to this value is similarly given by:
\beq
 k  \approx |y|^{-1} = \frac{q^2}{Q^2} = - a \frac{d(\ep/c_s)}{d H^{-1}} \label{k_final}
 \eeq

%\begin{figure}
%\begin{center}
%%\includegraphics[scale=0.5]{ns_epsilon.pdf}
%\includegraphics[scale=0.5]{rho_exit.pdf}
%\includegraphics[scale=0.5]{rho_star.pdf}
%\caption{ .}
%\label{figure_rho}
%\end{center}
%\end{figure}

We now evaluate the thermal fluctuations for the critical solution discussed in the main text. 
The model has a single free parameter, the 4-volume scale $B_0$. Its speed of sound is:
\beq\label{sound_crit1}
c_s^2\approx 2BX\approx \frac{4}{9}\epsilon^2B^2\rho^2\approx 
\frac{4}{9}\epsilon^2 (B_0\rho)^2\exp\left( 8\sqrt{ \frac{B_0 \rho}{3}}\right).
\eeq
%o find the speed of light profile we note that $1+w=2/3\epsilon=K/V\approx K/\rho$, so that 
%at leading order $K=\sqrt{X/B}$ and $X/B=4/9 \epsilon ^2 \rho^2$. Therefore
Conservation of entropy relates temperature to expansion and acoustic history:
\beq
g_0 T^3_0 = g_* \left( T a \over c_s\right)^3 \Rightarrow T_c = T_* a_* = T_0 \left(g_0 \over g_*\right)^{1/3},
\eeq
where 
\beq
g_0 = 3.91, T_0 = 2.73 ~{\rm K} = 2.35 \times 10^{-4} ~{\rm eV},
\eeq
are the effective number of degrees of freedom today, and the CMB temperature respectively. $T_*,a_*$, and $g_* \gtrsim 107$ are the temperature, scale factor, and the effective relativistic degrees of freedom at the end of the tachyacoustic phase, where $c^2_s \sim \frac{1}{3}, \ep \sim 2 \Rightarrow 6 B_0 \rho_* =  6 B_0 \times  \frac{\pi^2}{30}g_*T^4_* \sim 1$, using Eq. (\ref{sound_crit1}). Furthermore, Eq. (\ref{sound_crit1}) implies:
\beq
 \frac{d\ln (c_s/\ep) }{d\ln H} = -2\left(1+2\sqrt{\frac{B_0 \rho}{3}}\right). 
\eeq

Using Eq. (\ref{power_final})and Friedmann equation $3M^2_P H^2=\rho \propto a^{-2\ep}$ (assuming constant $\ep$) leads to an expression for the power spectrum:
\bea
{\cal P}_\zeta &=& \frac{d\ln(c_s/\ep)}{d H^{-1}} \frac{T_c}{2\pi^3 M^2_Pa }  \nonumber\\ &=&\frac{5^{1/4} 6^{\frac{1}{2\ep}}}{3^{1/2}\pi^{7/2}} g_*^{-1/4} \frac{(B_0\rho)^{\frac{1}{2}+ \frac{1}{2\ep}}}{(B_0 M^4_P)^{3/4}} \left(1+2\sqrt{\frac{B_0 \rho}{3}}\right),
\eea
and the sound horizon crossing wavenumber:
\bea
k=- a \frac{d(\ep/c_s)}{d H^{-1}}= \frac{3^{1/2} \pi^{1/2}}{6^{\frac{1}{2\ep}} 5^{1/4}} g_0^{1/3} g_*^{-1/12} \times \nonumber\\  \frac{T_0\left(1+2\sqrt{\frac{B_0 \rho}{3}}\right)\exp\left(-4\sqrt{\frac{B_0\rho}{3}}\right)}{(B_0 M_P^4)^{1/4} (B_0\rho)^{\frac{1}{2}+\frac{1}{2\ep}}},
\eea
which can be combined to give:
\beq\label{eq:B0rho}
\frac{{\cal P}_\zeta}{k^3} = \frac{5 \cdot 6^{2/\ep}}{9\pi^5 g_0} \frac{(B_0\rho)^{2+2/\ep}}{T^3_0} \frac{\exp\left(4\sqrt{3B_0\rho}\right)}{ \left(1+2\sqrt{\frac{B_0 \rho}{3}}\right)^2}.
\eeq

Subsequently, for scalar spectral index we get:
%\beq
%n_s-1= \frac{d\ln {\cal P}_s }{d \ln k} = - \frac{1+ \frac{2(1+2\ep)}{1+\ep}\sqrt{B_0 \rho \over 3}}{\left(1+2\sqrt{B_0 \rho \over 3} \right) \left(1+ \frac{4\ep}{1+\ep}\sqrt{B_0 \rho \over 3} \right)},
%\eeq
\bea
n_S-1= \frac{d\ln {\cal P}_\zeta }{d \ln k} = - \frac{1+ \frac{2(1+2\ep)}{1+\ep}\sqrt{B_0 \rho \over 3}}{1+\left( 2+4\epsilon \over 1+\ep\right) \sqrt{B_0 \rho \over 3} + \frac{8\ep B_0 \rho}{3(1+\ep)} } \nonumber \\ 
= -\frac{1+2\ep}{4\ep} \left(B_0 \rho \over 3\right)^{-1/2} + {\cal O} (B_0\rho)^{-1},  \label{ns_model1}
\eea
while its running is given by:
\beq
\frac{dn_S}{d \ln k} = -\frac{3}{16\ep}\frac{(1+2\ep)}{B_0 \rho} + {\cal O}\left(B_0 \rho\right)^{-3/2}. \label{run_model}
\eeq

Up to here, we have provided results for general $\ep$. However, we shall argue below that $\ep \rightarrow \infty$ is the only expected consistent asymptotic behavior as we approach the critical bimetric model in the UV limit, and thus we shall focus our predictions to $\ep \rightarrow \infty$.

Planck 2015 TT,TE,EE+lowP+lensing+ext (1-sigma) Table 4 \cite{Ade:2015xua} 
gives:
\bea
 {\cal P}_{\zeta}= (2.142 \pm 0.049) \times
10^{-9},  \\
{\rm at}~~ k = 0.05~{\rm Mpc}^{-1}  = 3.198 \times 10^{-31} ~{\rm eV}.
\eea

For these values, Eq. (\ref{eq:B0rho}) can be solved iteratively to give:
\beq
B_0\rho = \frac{\rho}{6\rho_*}\approx 580 {\rm ~~for~~} \ep=\infty
\eeq 

We see that the spectral index as $\ep \rightarrow \infty$ becomes:
\beq
n_S(\ep=\infty)  \approx 0.96478(64). \label{ns_predict}
\eeq
The theoretical uncertainty is estimated by difference between the first and second lines in Eq. (\ref{ns_model1}), as the analytic model is only reliable to leading order in $(B_0\rho)^{-1/2}$. 

The prediction \ref{ns_predict} is well within the observed range (Table 4 in \cite{Ade:2015xua}):
\bea
&n_S = 0.9667 \pm 0.0040,& \label{planck_ns}  \\ &{\rm Planck~ 2015~+lowP+lensing+ext. }& \nonumber
\eea

For, $\ep \rightarrow \infty$ (and $g_* \approx 107 $), we can further find the density and temperature at the end of the tachyacoustic phase:
\bea
\frac{\rho_*}{M^4_P} &=& \frac{1}{6 B_0 M^4_P} = 9.0 \times 10^{-14}, \\
\Rightarrow T_* &=& 2.24 \times 10^{-4} M_P =  5.5 \times 10^{14}~ {\rm GeV}.
\eea
Eqs.  (\ref{ns_model1}-\ref{run_model}) also give the running for the scalar spectral index for $\ep \rightarrow \infty$:
\beq
\frac{dn_S}{d\ln k} = -\frac{3}{2} (n_S-1)^2 = - 1.8 \times 10^{-3},
\eeq
well within the allowed observational range of $(-6.5 \pm 7.6) \times 10^{-3}$ (Eq. 42 in \cite{Ade:2015xua}). 
  
\subsection{B. Approaching the UV limit, and $\ep \rightarrow \infty$}

While we have used symmetry and consistency principles to construct the UV limit of the critical bimetric theory in the main text,  the subleading UV behavior remains unconstrained, and is reflected by the freedom to choose $\epsilon$. Here, we carefully study this subleading behavior and prove that the only consistent choice is to have $\ep \rightarrow \infty$ as $c_s \rightarrow \infty$.  

Equations (3), (8), and (9) in the main text provide the action for the critical UV tachyacoustic model. Let us write the sub-leading corrections to the Lagrangian (setting $M_P=1$ for simplicity):
\beq
S_\phi = \int d^4x \sqrt{-g} \left[\frac{\sqrt{1+ B_0 \phi^2 (\partial\phi)^2}}{B_0\phi^2} - \frac{3(\ln\phi)^2}{4B_0}   - \frac{W(\phi)}{B_0^{3/2}} \right],
\eeq 
where we have extracted the pre-factor $B^{-3/2}_0$ for $W(\phi)$, as subleading corrections are suppressed by powers of $B^{-1/2}_0$. Note that the kinetic term is fully fixed by the $EAdS_2\times E_3$ symmetry in the no-gravity limit, and thus the subleading correction, $W(\phi)$, only appears in the potential. Now, combining Friedmann and homogenous field equations for this action, and expanding in powers of $B^{-1/2}_0$, we find:
\begin{widetext}
\beq
\frac{B_0^{-3/2}}{\dot{\phi}^3 \phi^4 \ln\phi}\left\{ \ddot{\phi}\phi\ln\phi + \dot{\phi}^2\left[1+3\ln\phi+\left(W(\phi)-W'(\phi)\phi \ln\phi\right)\phi^3\dot{\phi}\right] \right\} +{\cal O} (B_0^{-2}) =0. \label{field_sub}
\eeq
\end{widetext}
This equation is trivial at ${\cal O}(B^{-1}_0)$, demonstrating the nature of the UV cuscuton limit which allows for arbitrary expansion (or field) history at this order. However, at ${\cal O}(B^{-3/2}_0)$ we have dynamical field equations. 

For $W(\phi)$ to be subdominant at early times, it should drop (or grow more slowly than $\ln^2\phi$) at large $\phi$. If we assume a power-law asymptotic fall-off, i.e. $W(\phi) = W_0 \phi^{-n}$ with $n \geq 0$, Eq. (\ref{field_sub}) has a power-law asymptotic solution:
\beq
\phi(t) =\left[W_0 \over (n-4) t\right]^{\frac{1}{n-4}}. 
\eeq

Requiring 
\bea
c_s \propto \phi\dot{\phi} \propto t^{-\frac{n-6}{n-4}} \rightarrow \infty, \\
\rho \sim H^2 \propto (\ln\phi)^2 \propto \left(\ln t^{-\frac{1}{n-4}}\right)^2  \rightarrow \infty,
\eea
as $t\rightarrow 0$, implies $n >6$.  

Finally, 
\beq
\ep \equiv -\frac{\dot{H}}{H^2} \propto \frac{1}{t (\ln t)^2} \rightarrow \infty, {\rm ~as~} t\rightarrow 0.
\eeq
\centerline{\it Q.E.D.}

\end{document}